\documentclass[preprint]{revtex4}
\usepackage{graphicx}
\usepackage{hyperref}
\usepackage{natbib}
\usepackage{hypernat}
\usepackage{amsmath}

\newcommand{\point}[1]{\ensuremath{{\rm #1}}}
\newcommand{\subP}{_\point{P}}
\newcommand{\Eqref}[1]{\eqref{eq:#1}}
\newcommand{\Eq}[1]{Eq.~\Eqref{#1}}

\newcommand{\ie}{{\em i.e.,\ }}
\newcommand{\eg}{{\em e.g.,\ }}

\begin{document}

\title{Spectral properties of the nonspherically decaying radiation
generated by a rotating superluminal source}

\author{Houshang Ardavan}
\affiliation{University of Cambridge}
\author{Arzhang Ardavan}
\affiliation{University of Oxford}
\author{John Singleton}
\email{jsingle@lanl.gov}
\author{Joseph Fasel}
\author{Andrea Schmidt}
\affiliation{Los Alamos National Laboratory}

\begin{abstract}
The focusing of the radiation
generated by a polarization current
with a superluminally rotating distribution pattern
is of a higher order in the plane of rotation
than in other directions.
Consequently,
our previously published asymptotic approximation
to the value of this field outside the equatorial plane
breaks down as the line of sight approaches a direction normal to the rotation axis,
\ie
is nonuniform with respect to the polar angle.
Here we employ an alternative asymptotic expansion
to show that,
though having a rate of decay with frequency ($\mu$)
that is by a factor of order $\mu^{2/3}$ slower,
the equatorial radiation field has the same dependence on distance
as the nonspherically decaying component of the generated field in other directions:
it, too, diminishes as the inverse square root of the distance from its source.
We also briefly discuss the relevance of these results
to the giant pulses received from pulsars:
the focused, nonspherically decaying pulses
that arise from a superluminal polarization current in a highly magnetized plasma
have a power-law spectrum
(\ie a flux density $S\propto\mu^\alpha$)
whose index ($\alpha$) is given by one of the values $-2/3$, $-2$, $-8/3$, or $-4$.

\end{abstract}

\maketitle

\section{Introduction}
Radiation by polarization currents
whose distribution patterns move faster than light {\it in vacuo}
has been the subject of several theoretical and experimental studies in recent years
\cite{BessarabAV:FasEsi,ArdavanA:Exponr,BessarabAV:Expser,BolotovskiiBM:Radssv,%
BolotovskiiBM:Radbcm,ArdavanH:Genfnd,ArdavanH:Speapc,ArdavanH:Morph}.
When the motion of its source is accelerated,
this radiation exhibits features that are not shared by any other known emission.
In particular,
the radiation from a rotating superluminal source
consists, in certain directions,
of a collection of subbeams
whose azimuthal and polar widths narrow
(as $R\subP^{-3}$ and $R\subP^{-1}$, respectively)
with distance $R\subP$ from the source
\cite{ArdavanH:Morph}.
Being composed of tightly focused wave packets
that are constantly dispersed and reconstructed out of other waves,
these subbeams neither diffract nor decay
in the same way as conventional radiation beams.
The field strength within each subbeam
diminishes as $R\subP^{-1/2}$,
instead of $R\subP^{-1}$,
with increasing $R\subP$
\cite{ArdavanH:Genfnd,ArdavanH:Speapc,ArdavanH:Morph}.

In earlier treatments
\cite{ArdavanH:Speapc,ArdavanH:Morph},
we evaluated the field of a superluminally rotating extended source
by superposing the fields of its constituent volume elements,
\ie
by convolving its density
with the familiar Li\'enard-Wiechert field of a rotating point source.
This Li\'enard-Wiechert field
is described by an expression essentially
identical to that which is encountered in the analysis of synchrotron radiation,
except that its value at any given observation time
receives contributions from more than one retarded time.
The multivalued nature of the retarded time
gives rise to the formation of caustics.
The wave fronts
emitted by each constituent volume element
of a superluminally moving accelerated source
possess a cusped envelope
on which the field is infinitely strong
(see Figs.~1 and 4 of Ref.~\cite{ArdavanH:Morph}).
Correspondingly,
the Green's function for the problem
is nonintegrably singular
for those source elements that approach the observer
along the radiation direction
with the speed of light and zero acceleration at the retarded time
(see Fig.~3 of Ref.~\cite{ArdavanH:Morph}):
the cusp of the envelope of wave fronts emanating from each such element
is a spiraling curve extending into the far zone
that passes through the position of the observer.
When the source oscillates at the same time as rotating,
the Hadamard finite part of the divergent integral
that results from convolving the Green's function with the source density
has a rapidly oscillating kernel for a far-field observation point.
The stationary points of the phase of this kernel
turn out to have different orders
depending on whether the observer is located in or out of the equatorial plane.

To reduce the complications
posed by the higher-order stationary points of this phase,
we restricted the asymptotic evaluation
of the radiation integral thus obtained
in Refs.~\cite{ArdavanH:Speapc,ArdavanH:Morph}
to observation points outside the plane of rotation,
\ie to spherical polar angles $\theta\subP$ that do not equal $\pi/2$.
The purpose of this paper
is to evaluate the field of a superluminally rotating extended source
also for the smaller class of observers
at polar coordinate $\theta\subP=\pi/2$
and to obtain, thereby,
a more global description of the nonspherically decaying radiation
that is generated by such a source.
The asymptotic expansion presented in Refs.~\cite{ArdavanH:Speapc,ArdavanH:Morph}
breaks down in the case of a subbeam
that is perpendicular to the rotation axis
because there is a higher-order focusing
associated with the waves emitted
by those source elements whose actual speeds
(rather than the line-of-sight components of their speeds)
equal the speed of light as they approach the observer with zero acceleration.

Here,
we present a brief account of the background material
on the radiation field of a rotating superluminal source in Section 2,
and the asymptotic evaluation of this field for an equatorial observer in Section 3.
In Section 4,
we give a description of the spectral properties
of the nonspherically decaying component of this radiation
in the light of the present results
and those obtained in Refs.~\cite{ArdavanH:Speapc,ArdavanH:Morph},
and discuss the relevance of these properties to pulsar observations.

\section{Background: radiation field of a rotating superluminal source}
We base our analysis
on a generic superluminal source distribution
\cite{ArdavanH:Speapc,ArdavanH:Morph},
which has been created in the laboratory
\cite{ArdavanA:Exponr}.
This source comprises a polarization current density
${\bf j}=\partial{\bf P}/\partial t$
for which
\begin{equation}
P_{r,\varphi,z}(r,\varphi,z,t)=
s_{r,\varphi,z}(r,z)\cos(m{\hat\varphi})\cos(\Omega t),\qquad -\pi<{\hat\varphi}\le\pi,
\label{eq:1}
\end{equation}
with
\begin{equation}
{\hat\varphi}\equiv\varphi-\omega t,
\label{eq:2}
\end{equation}
where $P_{r,\varphi,z}$
are the components of the polarization ${\bf P}$
in a cylindrical coordinate system based on the axis of rotation,
${\bf s}(r,z)$ is an arbitrary vector
that vanishes outside a finite region of the $(r,z)$ space,
and $m$ is a positive integer.
For fixed $t$,
the azimuthal dependence of the density \Eqref{1}
along each circle of radius $r$ within the source
is the same as that of a sinusoidal wave train,
of wavelength $2\pi r/m$,
whose $m$ cycles fit around the circumference of the circle smoothly.
As time elapses,
this wave train both propagates around each circle with the velocity $r\omega$
and oscillates in its amplitude with the frequency $\Omega$.
This is a generic source:
one can construct any distribution
with a uniformly rotating pattern
$P_{r,\varphi,z}(r,{\hat\varphi},z)$
by the superposition over $m$
of terms of the form $s_{r,\varphi,z}(r,z,m)\cos(m{\hat\varphi})$.

The electromagnetic fields
\begin{equation}
{\bf E}=
-{\bf\nabla}\subP A^0 - \partial{\bf A}/\partial(c t\subP),\quad
{\bf B}={\bf\nabla}\subP{\bf\times A}
\label{eq:3}
\end{equation}
arising from such a source are given,
in the absence of boundaries,
by the following classical expression for the retarded four-potential:
\begin{equation}
A^\mu({\bf x}\subP,t\subP)=
c^{-1}\int{\rm d}^3 x{\rm d}t\, j^\mu({\bf x},t)\delta(t\subP-t-R/c)/R,\quad \mu=0,\cdots,3.
\label{eq:4}
\end{equation}
Here,
$({\bf x}\subP,t\subP)=(r\subP,\varphi\subP,z\subP,t\subP)$
and $({\bf x},t)=(r,\varphi,z,t)$
are the space-time coordinates of the observation point and the source points, respectively,
$R$ stands for the magnitude of ${\bf R}\equiv{\bf x}\subP-{\bf x}$,
and $\mu=1,2,3$ designate the spatial components, ${\bf A}$ and ${\bf j}$,
of $A^\mu$ and $j^\mu$ in a Cartesian coordinate system.

To find the retarded field
that follows from \Eq{4}
for the source described in \Eq{1},
we first calculated in Ref.~\cite{ArdavanH:Speapc}
the Li\'enard-Wiechert field
arising from a circularly moving point source
with a speed $r\omega>c$,
\ie
a generalization of the synchrotron radiation
to the superluminal regime.
We then evaluated the integral
representing the retarded field of the extended source \Eqref{1}
by superposing the fields
generated by the constituent volume elements of this source,
\ie
by using the generalization of the synchrotron field
as the Green's function for the problem.
In the superluminal regime,
this Green's function has extended singularities
arising from the constructive intereference of the emitted waves
on the envelope of wave fronts and its cusp.

Labeling each element of the extended source \Eqref{1}
by its Lagrangian coordinate ${\hat\varphi}$
and performing the integration with respect to $t$ and ${\hat\varphi}$
(or equivalently $\varphi$ and ${\hat\varphi}$)
in the multiple integral implied by Eqs.~\Eqref{1}--\Eqref{4},
we showed in Ref.~\cite{ArdavanH:Speapc}
that the resulting expression for the radiation field
${\bf B}$ (or ${\bf E}$)
consists of two parts:
one that decays spherically
(as $R\subP^{-1}$, as in a conventional radiation field)
and another, ${\bf B}^{\rm ns}$
(with ${\bf E}^{\rm ns}={\hat{\bf n}}{\bf\times}{\bf B}^{\rm ns}$),
that decays nonspherically
(as $R\subP^{-1/2}$)
within the conical shell $\arcsin(1/{\hat r}_>)\le\theta\subP\le\arcsin(1/{\hat r}_<)$
in the far zone.
Here, $(R\subP,\theta\subP,\varphi\subP)$
are the spherical polar coordinates of the observation point $\point{P}$,
${\hat r}$ stands for $r\omega/c$,
${\hat{\bf n}}\equiv{\bf R}/R$ is a unit vector in the radiation direction,
and ${\hat r}_<>1$ and ${\hat r}_>>{\hat r}_<$
denote the radial boundaries of the support of the source density ${\bf s}$.

The expression for the nonspherically decaying component of the field
within this conical shell,
in the far zone,
is
\begin{equation}\begin{split}
{\bf B}^{\rm ns}\simeq
&-\textstyle{4\over3}{\rm i}\exp[{\rm i}(\Omega/\omega)(\varphi\subP+3\pi/2)]
\sum_{\mu=\mu_\pm}\mu\exp(-{\rm i}\mu{\hat\varphi}\subP)\\
&\times\sum_{j=1}^3 {\bar q}_j\int_{\Delta\ge 0}{\hat r}{\rm d}{\hat r}\,{\rm d}{\hat z}\,
\Delta^{-1/2}{\bf u}_j\exp(-{\rm i}\mu\phi_-),
\end{split}\label{eq:5}
\end{equation}
where $\mu_\pm\equiv(\Omega/\omega)\pm m$,
${\hat\varphi}\subP\equiv\varphi\subP-\omega t\subP$,
\begin{equation}
{\bar q}_j\equiv(1\qquad-{\rm i}\Omega/\omega\qquad{\rm i}\Omega/\omega),
\label{eq:6}
\end{equation}
\begin{equation}
{\bf u}_1\equiv
s_r\cos\theta\subP{\hat{\bf e}}_\parallel+s_\varphi {\hat{\bf e}}_\perp,\quad
{\bf u}_2\equiv -s_\varphi\cos\theta\subP{\hat{\bf e}}_\parallel+s_r{\hat{\bf e}}_\perp,\quad
{\bf u}_3\equiv-s_z\sin\theta\subP{\hat{\bf e}}_\parallel,
\label{eq:7}
\end{equation}
\begin{equation}
\Delta\equiv({\hat r}\subP^2-1)({\hat r}^2-1)-({\hat z}-{\hat z}\subP)^2,
\label{eq:8}
\end{equation}
\begin{equation}
\phi_\pm\equiv{\hat R}_\pm+\varphi_\pm-\varphi\subP,
\label{eq:9}
\end{equation}
\begin{equation}
\varphi_\pm=\varphi\subP+2\pi-\arccos[(1\mp\Delta^{1/2})/({\hat r}{\hat r}\subP)],
\label{eq:10}
\end{equation}
and
\begin{equation}
{\hat R}_\pm\equiv
[({\hat z}-{\hat z}\subP)^2+{\hat r}^2+{\hat r}\subP^2-2(1\mp\Delta^{1/2})]^{1/2};
\label{eq:11}
\end{equation}
see Eq.~(47) of Ref.~\cite{ArdavanH:Speapc}.  In this expression,
$({\hat r},{\hat z};{\hat r}\subP,{\hat z}\subP)$
stand for $(r\omega/c, z\omega/c;r\subP\omega/c, z\subP\omega/c)$,
and ${\hat{\bf e}}_\parallel\equiv
{\hat{\bf e}}_z\times{\hat{\bf n}}/|{\hat{\bf e}}_z\times{\hat{\bf n}}|$
(which is parallel to the plane of rotation)
and ${\hat{\bf e}}_\perp\equiv{\hat{\bf n}}{\bf\times}{\hat{\bf e}}_\parallel$
comprise a pair of unit vectors normal to the radiation direction ${\hat{\bf n}}$
(${\hat{\bf e}}_z$ is the base vector associated with the coordinate $z$).
The domain of integration
consists of the part of the source distribution ${\bf s}(r,z)$
that falls within $\Delta\ge0$
(see Fig.~4 of Ref.~\cite{ArdavanH:Morph}).

Both derivatives,
$\partial\phi_-/\partial {\hat r}$ and $\partial\phi_-/\partial {\hat z}$,
of the function $\phi_-({\hat r},{\hat z})$
that appears in the phase of the integrand in \Eq{5}
vanish at the point ${\hat r}=1$, ${\hat z}={\hat z}\subP$,
where the cusp curve of the bifurcation surface
is tangent to the light cylinder
(see Figs.~3 and 4 of Ref.~\cite{ArdavanH:Morph}).
However,
$\partial^2\phi_-/\partial{\hat r}^2$ diverges at this point,
so that neither the phase nor the amplitude
of the kernel of the integral in \Eq{5}
are analytic at ${\hat r}=1,{\hat z}={\hat z}\subP$.
Only for an observer who is located outside the plane of rotation,
\ie
whose coordinate $z\subP$ does not match the coordinate $z$ of any source element,
is the function $\phi_-(r,z)$ analytic throughout the domain of integration.
To take advantage of the simplifications
offered by the analyticity of $\phi_-$ as a function of $r$,
we restricted the analyses in Refs.~\cite{ArdavanH:Speapc,ArdavanH:Morph}
to observation points for which $\theta\subP\ne\pi/2$.

In the calculation that follows,
we find an asymptotic approximation to the integral
\begin{equation}
{\cal I}\equiv
\int_{\Delta\ge 0}{\hat r}{\rm d}{\hat r}\,{\rm d}{\hat z}\,\Delta^{-1/2}{\bf u}_j
\exp(-{\rm i}\mu\phi_-)
\label{eq:12}
\end{equation}
in \Eq{5}
that is valid in the plane of rotation,
\ie for $\theta\subP=\pi/2$.
We shall treat only the case of positive $\mu$;
${\cal I}(\mu)$ for negative $\mu$ can then be obtained
via ${\cal I}(-\mu)={\cal I}(\mu)^*$.

\section{Asymptotic value of the field for an equatorial observer in the far zone}
Since the main contribution
toward the value of the field at $\theta\subP=\pi/2$
is made by the source elements
that lie in the vicinity of the critical point
${\hat r}=1$, ${\hat z}={\hat z}\subP$,
the first step in the asymptotic evaluation of ${\bf B}^{\rm ns}$
is to replace $({\hat r},{\hat z})$
by a new pair of variables $(\rho,\sigma)$
for which the phase function $\phi_-(\rho,\sigma)$
is rendered analytic at this point:
\begin{equation}
{\hat r}=(1+\rho^2\cosh^2\sigma)^{1/2},
\label{eq:13}
\end{equation}
\begin{equation}
{\hat z}={\hat z}\subP+({\hat r}\subP^2-1)^{1/2}\rho\sinh\sigma
\label{eq:14}
\end{equation}
This transformation
replaces ${\hat r}\Delta^{-1/2}{\rm d}{\hat r}{\rm d}{\hat z}$
by $\rho\cosh\sigma{\rm d}\rho{\rm d}\sigma$
and yields
\begin{equation}\begin{split}
\phi_-(\rho,\sigma)=
&[{\hat r}\subP^2-1-2({\hat r}\subP^2-1)^{1/2}\rho
+({\hat r}\subP^2\sinh^2\sigma+1)\rho^2]^{1/2}+2\pi\\
&-\arccos\{{\hat r}\subP^{-1}(1+\rho^2\cosh^2\sigma)^{-1/2}[1+({\hat r}\subP^2-1)^{1/2}\rho]\},
\end{split}\label{eq:15}
\end{equation}
which is analytic at $\rho=\sigma=0$.
In the plane of rotation,
\ie for $\sigma=0$,
the two critical points
designated as $\point{C}$ and $\point{S}$ in Refs.~\cite{ArdavanH:Speapc,ArdavanH:Morph}
coalesce,
and both derivatives,
$\partial\phi_-/\partial\rho$ and $\partial^2\phi_-/\partial\rho^2$,
of the resulting function $\phi_-(\rho,0)$ vanish at $\rho=0$
[see Eqs.~\Eqref{15} and \Eqref{24} below],
so that the function $\phi_-(\rho,\sigma)$ is stationary at $\rho=\sigma=0$.

To see that applying the method of stationary phase to the integral in \Eq{5}
results in a valid asymptotic approximation for large ${\hat R}\subP$,
let us begin by casting the $\sigma$ dependence of the phase $\phi_-$
into a canonical form
\cite{BorvikovVA:UnStaPha}.
Since $\sigma=0$ is an isolated stationary point of $\phi_-$
(when regarded as a function of the single variable $\sigma$),
we may employ the following transformation:
\begin{equation}
\phi_-=\phi_-|_{\sigma=0}+\textstyle{1\over2}b\zeta^2,
\label{eq:16}
\end{equation}
in which
\begin{equation}
b\equiv{\partial^2\phi_-\over\partial\sigma^2}\Big|_{\sigma=0}=
{\rho^2[{\hat r}\subP^2-1-({\hat r}\subP^2-1)^{1/2}\rho+{\hat r}\subP^2\rho^2]
\over(1+\rho^2)[({\hat r}\subP^2-1)^{1/2}-\rho]}.
\label{eq:17}
\end{equation}
Equation \Eqref{16} expresses $\sigma$ as a function of $\zeta$ implicitly.
Repeated differentiations of this equation with respect to $\zeta$
result in
\begin{equation}
{\partial\phi_-\over\partial\sigma}{\partial\sigma\over\partial\zeta}=b\zeta,
\label{eq:18}
\end{equation}
\begin{equation}
{\partial\phi_-\over\partial\sigma}{\partial^2\sigma\over\partial\zeta^2}
+{\partial^2\phi_-\over\partial\sigma^2}\Big({\partial\sigma\over\partial\zeta}\Big)^2
=b,
\label{eq:19}
\end{equation}
and so on,
which when evaluated at $\zeta=0$
supply the coefficients
$\partial\sigma/\partial\zeta|_{\sigma=0}$,
$\partial^2\sigma/\partial\zeta^2|_{\sigma=0}$,
etc.,
in the Taylor expansion of $\sigma$
in powers of $\zeta$.

The integral ${\cal I}$ in \Eq{12}
can therefore be written as
\begin{equation}
{\cal I}=\int{\rm d}\rho{\rm d}\zeta Q(\rho,\zeta)\exp(-{\rm i}\beta\zeta^2),
\label{eq:20}
\end{equation}
where
\begin{equation}
Q(\rho,\zeta)=
\rho\cosh\sigma{\bf u}_j\exp(-{\rm i}\mu\phi_-|_{\sigma=0})\partial\sigma/\partial\zeta,
\label{eq:21}
\end{equation}
\begin{equation}
{\partial\sigma\over\partial\zeta}=
{b\zeta{\hat R}_-(1+\rho^2\cosh^2\sigma)\over\rho^2\sinh\sigma\cosh\sigma}
\big[{\hat r}\subP^2-1-({\hat r}\subP^2-1)^{1/2}\rho
+{\hat r}\subP^2\rho^2\cosh^2\sigma\big]^{-1},
\label{eq:22}
\end{equation}
and $\beta\equiv\textstyle{1\over2}\mu b$.
The limits of integration
are determined by the image of $\Delta\ge0$ under transformation \Eqref{16}.

The parameter $b$ that multiplies the phase of the integrand in \Eq{20}
has a large value in the far zone:
\begin{equation}
b\simeq\rho^2{\hat r}\subP,\qquad{\hat R}\subP\gg1,
\label{eq:23}
\end{equation}
[see \Eq{17}].
The asymptotic value of the integral ${\cal I}$ for large ${\hat R}\subP$
therefore receives its leading contribution
from the immediate vicinity of $\zeta=0$,
where the phase of its integrand is stationary
\cite{BorvikovVA:UnStaPha}.
Replacing $Q(\rho,\zeta)$ in \Eq{20} by $Q(\rho,0)$
and extending the range of integration with respect to $\zeta$ to $(-\infty,\infty)$,
we obtain
\begin{equation}\begin{split}
{\cal I}
&\simeq(2\pi/\mu)^{1/2}
\int{\rm d}\rho\,
\rho{\bf u}_j|_{\sigma=0}b^{-1/2}
\exp[-{\rm i}(\mu\phi_-|_{\sigma=0}+\pi/4)](\partial\sigma/\partial\zeta)_{\sigma=0}\\
&\simeq(2\pi/\mu)^{1/2}{\hat r}\subP^{-1/2}\exp\{-{\rm i}[\mu({\hat r}\subP+3\pi/2)+\pi/4]\}\\
&\quad\times\int{\rm d}\rho\,{\bf u}_j|_{{\hat r}=(1+\rho^2)^{1/2},{\hat z}={\hat z}\subP}
\exp[-{\rm i}\mu(\arctan\rho-\rho)],\qquad
{\hat R}\subP\gg1,
\end{split}\label{eq:24}
\end{equation}
where the integation extends over all values of $\rho$
for which the source density
${\bf s}|_{{\hat r}=(1+\rho^2)^{1/2},{\hat z}={\hat z}\subP}$
is nonzero
[see \Eq{7}].
In deriving this expression,
we have inferred the value $\partial\sigma/\partial\zeta|_{\sigma=0}=1$
of the indeterminate Jacobian
that appears in $Q(\rho,0)$
from \Eq{19}
[or, equivalently, from \Eq{22} and l'H\^opital's rule],
and expressed $\phi_-|_{\sigma=0}$ and $b$
in terms of their far-field values
by means of Eqs.~\Eqref{15} and \Eqref{23}.

The contribution ${\bf B}^{\rm ns}$
toward the magnetic field ${\bf B}$ of the radiation
is made by those volume elements of the source
that approach the observation point $\point{P}$,
along the radiation direction,
with the speed of light and zero acceleration at the retarded time,
\ie by the source elements for which $\Delta=0$.
Hence,
the amplitude of the integrand in \Eq{5}
has already been approximated by its leading term
in powers of $\Delta^{1/2}=({\hat r}\subP^2-1)^{1/2}\rho$
(see Refs.~\cite{ArdavanH:Speapc,ArdavanH:Morph}).
To be consistent,
we must also approximate the amplitude of the integrand in \Eq{24}
by its value for $\rho\ll1$:
\begin{equation}\begin{split}
{\cal I}\simeq
&(2\pi/\mu)^{1/2}{\hat r}\subP^{-1/2}{\bf u}_j|_{{\hat r}=1,{\hat z}={\hat z}\subP}
\exp\{-{\rm i}[\mu({\hat r}\subP+3\pi/2)+\pi/4]\}\\
&\times\int_0^{({\hat r}_>^2-1)^{1/2}}{\rm d}\rho\,\exp[-{\rm i}\mu(\arctan\rho-\rho)],
\end{split}\label{eq:25}
\end{equation}
where ${\hat r}_>$ denotes
the radial extent of the support of the source density ${\bf s}$.
This reduces to
\begin{equation}
{\cal I}=
3^{-2/3}\Gamma({\textstyle{1\over3}})(2\pi)^{1/2}\mu^{-5/6}{\bf u}_j
|_{{\hat r}=1,{\hat z}={\hat z}\subP}
\exp\{-{\rm i}[\mu({\hat r}\subP+3\pi/2)+\pi/12]\}{\hat r}\subP^{-1/2}
\label{eq:26}
\end{equation}
in the regime $\mu\gg1$,
where we can approximate $\arctan\rho-\rho$ in the argument of the exponential
by $-\textstyle{1\over3}\rho^3$
and replace the upper limit of integration by $\infty$
\cite{BorvikovVA:UnStaPha}.

Thus,
Eqs.~\Eqref{5}, \Eqref{12} and \Eqref{25} jointly yield
the following expression for the leading term in the asymptotic expansion,
for ${\hat R}\subP\gg1$,
of the magnetic field of the radiation
close to the plane $\theta\subP=\pi/2$:
\begin{equation}\begin{split}
{\bf B}\simeq
&-{4\over3}{\rm i}(2\pi)^{1/2}{\hat R}\subP^{-1/2}\csc^{1/2}\theta\subP
\exp[{\rm i}(\Omega/\omega)(\varphi\subP+3\pi/2)]\\
&\times\sum_{\mu=\mu_\pm}|\mu|^{1/2}{\rm sgn}(\mu)
\exp\{-{\rm i}[\mu({\hat\varphi}\subP+{\hat r}\subP+3\pi/2)+{\pi\over4}{\rm sgn}(\mu)]\}\\
&\times\sum_{j=1}^3{\bar q}_j{\bf u}_j\big|_{{\hat r}=1,{\hat z}={\hat z}\subP}{\cal J},
\end{split}\label{eq:27}
\end{equation}
where
\begin{equation}
{\cal J}\equiv\int_0^{({\hat r}_>^2-1)^{1/2}}{\rm d}\rho\,\exp[-{\rm i}\mu(\arctan\rho-\rho)];
\label{eq:28}
\end{equation}
for,
the contribution ${\bf B}^{\rm ns}$
toward the magnetic field ${\bf B}$ of the radiation
is larger by a factor of the order of ${\hat R}\subP^{1/2}$
than the spherically decaying contribution.
This is the counterpart of Eq.~(55) of Ref.~\cite{ArdavanH:Speapc}
and Eq.~(61) of Ref.\cite{ArdavanH:Morph}
(the electric-field vector of this radiation
is given by ${\hat{\bf n}}{\bf\times B}$ as in any other radiation).

Note that the remaining integral in the above expression
reduces to
\begin{equation}
{\cal J}\simeq3^{-2/3}\Gamma({\textstyle{1\over3}})\exp({\rm i}\pi/6)\mu^{-1/3}
\label{eq:29}
\end{equation}
in the limit $\vert\mu\vert\gg1$ [see \Eq{26}].

\section{Spectrum of the nonspherically decaying radiation: relevance to pulsar observations}
\Eq{27} shows
that the radiation field of a rotating superluminal source
diminishes as ${\hat R}\subP^{-1/2}$
with the distance ${\hat R}\subP$
also in the equatorial plane $\theta\subP=\pi/2$.
This differs from the corresponding result for $\theta\subP\neq\pi/2$
[Eq.~(55) of Ref.~7]
mainly in its dependence on frequency.
The Fourier transform ${\bar{\bf s}}$ in Eq.~(57) of Ref.~\cite{ArdavanH:Speapc}
has the asymptotic dependence $\mu^{-1}$ on $\mu$
for a source density ${\bf s}(r,z)$ that is a smooth function of $z$.
Therefore,
when ${\bf s}(z)$ is smooth
and the radiation frequency $\vert\mu\omega\vert$
appreciably exceeds the rotation frequency $\omega$,
the field in the plane of rotation decays more slowly with frequency,
by a factor of order $\mu^{2/3}$,
than does the field outside this plane.

Since the azimuthal width of the generated subbeams
(and hence the duration of the narrow signals that constitute the overall pulse)
is independent of frequency,
the flux density $S$ of such signals
(\ie the power propagating across a unit area per unit frequency)
is proportional to $\vert{\bf B}\vert^2/\Delta\mu$,
where $\Delta\mu\sim\vert\mu\vert$ is the bandwidth of the radiation.
The flux density of the emission described by \Eq{27}
thus depends on frequency as $S\propto\mu^{-2/3}{{\bar q}_j}^2\vert{\bf s}(\mu)\vert^2$
for $\vert\mu\vert\gg1$.
Here,
$\vert{\bf s}(\mu)\vert$ designates
the frequency dependence of the factor ${\bf s}$
that enters the expression for the polarization ${\bf P}$
and the definitions of ${\bf u}_j$
[see Eqs.~\Eqref{1} and \Eqref{7}].
The flux density of the corresponding emission outside the equatorial plane
depends on frequency as $S\propto\mu^{-2}{{\bar q}_j}^2\vert{\bf s}(\mu)\vert^2$
for $\vert\mu\vert\gg1$
since,
apart from the dependence ${\bf s}(\mu)$ of a mutiplicative factor
(such as electric susceptibility)
in ${\bf s}$,
the Fourier transform ${\bar{\bf s}}$ would decay as $\mu^{-1}$ for a smooth ${\bf s}(z)$.

To compare the predictions of \Eq{27} [and Eq.~(55) of Ref.~7]
with the observed spectra of the giant pulses from pulsars,
we therefore need to estimate the frequency dependence of the electric susceptibility
(contained in the factor ${\bf s}$)
for the magnetospheric plasma of a pulsar.
The simple classical model
of propagation of electromagnetic disturbances in a cold magnetized plasma
yields a dielectric tensor,
and hence an electric susceptibility,
whose components decay with frequency as $(\mu\omega)^{-1}$
when the frequency $\mu\omega$ of the disturbance that polarizes the medium
is much lower than the gyration frequency of the electrons in the magnetized plasma;
see, \eg Eq.~(7.67) of \cite{JacksonJD:Classical}.
For a magnetic field as strong as that of a pulsar
($\sim 10^{12}$ G),
the Larmor frequency of an electron exceeds the highest radio frequencies
at which the pulses are observed
by a factor of order $10^6$,
so that ${\bf s}(\mu)\propto\mu^{-1}$ for pulsars.

Using this result,
we obtain $S\propto{{\bar q}_j}^2\mu^{-8/3}$ for $\theta\subP=\pi/2$,
and $S\propto{{\bar q}_j}^2\mu^{-4}$ for $\theta\subP\neq\pi/2$.
Depending on whether the modulation frequency $\Omega$
in the expression for ${\bar q}_j$ [\Eq{6}]
is comparable to or much smaller than the frequency $m\omega$
of the sinusoidal wave train characterizing the spatial distribution of the source,
therefore,
the spectral density of the nonspherically-decaying radiation is given by
\begin{equation}
S\propto\mu^{-2/3},\qquad\theta\subP=\pi/2,\quad\Omega/\omega\simeq\vert\mu\vert,
\label{eq:28a}
\end{equation}
\begin{equation}
S\propto\mu^{-2},\qquad\theta\subP\neq\pi/2,\quad\Omega/\omega\simeq\vert\mu\vert,
\label{eq:28b}
\end{equation}
\begin{equation}
S\propto\mu^{-8/3},\qquad\theta\subP=\pi/2;
\quad\Omega/\omega\ll\vert\mu\vert\quad{\rm or}\quad j=1,
\label{eq:28c}
\end{equation}
or
\begin{equation}
S\propto\mu^{-4},\qquad\theta\subP\neq\pi/2;
\quad\Omega/\omega\ll\vert\mu\vert\quad{\rm or}\quad j=1.
\label{eq:28d}
\end{equation}
In other words,
the spectral index of the pulses portraying the subbeams
can have any of the values $-2/3$, $-2$, $-8/3$, or $-4$.

The range of spectral indices
($-4\leq\alpha\leq-2/3$)
implied by \Eq{27} and its counterpart,
Eq.~(57) of Ref.~\cite{ArdavanH:Speapc},
is consistent
with that which characterizes the observed power-law spectra
of the giant pulses from pulsars
\cite{SallmenS:Simdog,KinkhabwalaA:Multi,PopovMV:Instant}.
For radio pulsars,
the rotation frequency $\omega$
of the distribution pattern of the radiating polarization current
is of the order of $1$ rad/sec,
and the oscillation frequency $\mu\omega/2\pi$ of the source density,
of the order of $100$ MHz,
so that $\mu$ has a large value of the order of $10^9$.
The coherent component of the radiation,
\ie
the sharply focused subbeams
that decay as ${\hat R}\subP^{-1/2}$,
are emitted at the frequency $\mu\omega$
\cite{ArdavanH:Speapc}.
The spherically decaying, incoherent component
of the radiation arising from the polarization current described in \Eq{1},
on the other hand,
contains frequencies
that are higher than $\mu\omega$ by a factor of order $(\Omega/\omega)^2$
\cite{ArdavanH:Fresfb}.
In pulsars,
$(\Omega/\omega)^2\sim10^{18}$ when $\Omega/\omega$ is comparable to $\mu$,
\ie when the frequency $m\omega$
that characterizes the spatial fluctuations of the emitting plasma
is of the order of, or smaller than, its modulation frequency $\Omega$.
Hence,
not only the power-law indices of the coherent component,
but the unusually broad spectral distribution
of the incoherent component of this radiation, too,
is consistent with the observational data from certain pulsars.
The pulsed emission from the Crab pulsar,
for example,
extends over 53 octaves of the electromagnetic spectrum
from radio waves to $\gamma$-rays
\cite{LorimerD:HbPA}.

We note, finally, that neither the asymptotic expansion presented here
nor that which was obtained in Refs.~\cite{ArdavanH:Speapc,ArdavanH:Morph}
are uniform with respect to the parameter $\theta\subP$.
The present approximation
receives contributions
only from the volume elements in the vicinity of the single source point
${\hat r}=1$, ${\hat z}={\hat z}\subP$
at which the cusp curve $\Delta=0$ of the bifurcation surface
touches the light cylinder
(see Figs.~3 and 4 of Ref.~\cite{ArdavanH:Speapc}).
This is in sharp contrast
to the asymptotic expansion of the field outside the equatorial plane,
where the leading term receives contributions
from a filamentary locus of source elements,
\ie
from the intersection of the cusp curve of the bifurcation surface
with the entire volume of the source.
Comparison of \Eq{27} with its counterpart,
Eq.~(57) of Ref.~\cite{ArdavanH:Speapc},
shows that the (smooth) transition from the nonequatorial to the equatorial regime
occurs across $\theta\subP\simeq\arccos(\mu^{-2/3})$.
However,
the derivation of a uniform asymptotic approximation to integral ${\cal I}$
that would determine the field in the transition region
is a challenging mathematical problem that remains open.

\section*{Acknowledgemens}
H.\ A.\ thanks Janusz Gil for helpful conversations.
A.\ A.\ is supported by the Royal Society.
J.\ S., J.\ F., and A.\ S.\ are supported
by U.S.\ Department of Energy grant LDRD 20050540ER.

\bibliography{jabosa,superluminal}

\begin{thebibliography}{15}
\expandafter\ifx\csname natexlab\endcsname\relax\def\natexlab#1{#1}\fi
\expandafter\ifx\csname bibnamefont\endcsname\relax
  \def\bibnamefont#1{#1}\fi
\expandafter\ifx\csname bibfnamefont\endcsname\relax
  \def\bibfnamefont#1{#1}\fi
\expandafter\ifx\csname citenamefont\endcsname\relax
  \def\citenamefont#1{#1}\fi
\expandafter\ifx\csname url\endcsname\relax
  \def\url#1{\texttt{#1}}\fi
\expandafter\ifx\csname urlprefix\endcsname\relax\def\urlprefix{URL }\fi
\providecommand{\bibinfo}[2]{#2}
\providecommand{\eprint}[2][]{\url{#2}}

\bibitem[{\citenamefont{Bessarab et~al.}(2004)\citenamefont{Bessarab, Gorbunov,
  Martynenko, and Prudkoy}}]{BessarabAV:FasEsi}
\bibinfo{author}{\bibfnamefont{A.~V.} \bibnamefont{Bessarab}},
  \bibinfo{author}{\bibfnamefont{A.~A.} \bibnamefont{Gorbunov}},
  \bibinfo{author}{\bibfnamefont{S.~P.} \bibnamefont{Martynenko}},
  \bibnamefont{and} \bibinfo{author}{\bibfnamefont{N.~A.}
  \bibnamefont{Prudkoy}}, \bibinfo{journal}{IEEE Trans. Plasma Sci.}
  \textbf{\bibinfo{volume}{32}}, \bibinfo{pages}{1400} (\bibinfo{year}{2004}),
  ISSN \bibinfo{issn}{0093-3813}.

\bibitem[{\citenamefont{Ardavan
  et~al.}(2004{\natexlab{a}})\citenamefont{Ardavan, Hayes, Singleton, Ardavan,
  Fopma, and Halliday}}]{ArdavanA:Exponr}
\bibinfo{author}{\bibfnamefont{A.}~\bibnamefont{Ardavan}},
  \bibinfo{author}{\bibfnamefont{W.}~\bibnamefont{Hayes}},
  \bibinfo{author}{\bibfnamefont{J.}~\bibnamefont{Singleton}},
  \bibinfo{author}{\bibfnamefont{H.}~\bibnamefont{Ardavan}},
  \bibinfo{author}{\bibfnamefont{J.}~\bibnamefont{Fopma}}, \bibnamefont{and}
  \bibinfo{author}{\bibfnamefont{D.}~\bibnamefont{Halliday}},
  \bibinfo{journal}{J. Appl. Phys.} \textbf{\bibinfo{volume}{96}},
  \bibinfo{pages}{7760} (\bibinfo{year}{2004}{\natexlab{a}}), ISSN
  \bibinfo{issn}{0021-8979}, \bibinfo{note}{corrected version of
  \textbf{96}(8), 4614--4631}.

\bibitem[{\citenamefont{Bessarab et~al.}(2006)\citenamefont{Bessarab,
  Martynenko, Prudkoi, Soldatov, and Terekhin}}]{BessarabAV:Expser}
\bibinfo{author}{\bibfnamefont{A.~V.} \bibnamefont{Bessarab}},
  \bibinfo{author}{\bibfnamefont{S.~P.} \bibnamefont{Martynenko}},
  \bibinfo{author}{\bibfnamefont{N.~A.} \bibnamefont{Prudkoi}},
  \bibinfo{author}{\bibfnamefont{A.~V.} \bibnamefont{Soldatov}},
  \bibnamefont{and} \bibinfo{author}{\bibfnamefont{V.~A.}
  \bibnamefont{Terekhin}}, \bibinfo{journal}{Radiation Physics and Chemistry}
  \textbf{\bibinfo{volume}{75}}, \bibinfo{pages}{825} (\bibinfo{year}{2006}),
  ISSN \bibinfo{issn}{0969-806X}.

\bibitem[{\citenamefont{Bolotovskii and Serov}(2006)}]{BolotovskiiBM:Radssv}
\bibinfo{author}{\bibfnamefont{B.~M.} \bibnamefont{Bolotovskii}}
  \bibnamefont{and} \bibinfo{author}{\bibfnamefont{A.~V.} \bibnamefont{Serov}},
  \bibinfo{journal}{Radiation Physics and Chemistry}
  \textbf{\bibinfo{volume}{75}}, \bibinfo{pages}{813} (\bibinfo{year}{2006}),
  ISSN \bibinfo{issn}{0969-806X}.

\bibitem[{\citenamefont{Bolotovskii and Bykov}(1990)}]{BolotovskiiBM:Radbcm}
\bibinfo{author}{\bibfnamefont{B.~M.} \bibnamefont{Bolotovskii}}
  \bibnamefont{and} \bibinfo{author}{\bibfnamefont{V.~P.} \bibnamefont{Bykov}},
  \bibinfo{journal}{Sov. Phys. Usp.} \textbf{\bibinfo{volume}{33}},
  \bibinfo{pages}{477} (\bibinfo{year}{1990}), ISSN \bibinfo{issn}{0038-5670}.

\bibitem[{\citenamefont{Ardavan}(1998)}]{ArdavanH:Genfnd}
\bibinfo{author}{\bibfnamefont{H.}~\bibnamefont{Ardavan}},
  \bibinfo{journal}{Phys. Rev. E} \textbf{\bibinfo{volume}{58}},
  \bibinfo{pages}{6659} (\bibinfo{year}{1998}), ISSN \bibinfo{issn}{1063-651X}.

\bibitem[{\citenamefont{Ardavan
  et~al.}(2004{\natexlab{b}})\citenamefont{Ardavan, Ardavan, and
  Singleton}}]{ArdavanH:Speapc}
\bibinfo{author}{\bibfnamefont{H.}~\bibnamefont{Ardavan}},
  \bibinfo{author}{\bibfnamefont{A.}~\bibnamefont{Ardavan}}, \bibnamefont{and}
  \bibinfo{author}{\bibfnamefont{J.}~\bibnamefont{Singleton}},
  \bibinfo{journal}{J. Opt. Soc. Am. A} \textbf{\bibinfo{volume}{21}},
  \bibinfo{pages}{858} (\bibinfo{year}{2004}{\natexlab{b}}), ISSN
  \bibinfo{issn}{1084-7529}.

\bibitem[{\citenamefont{Ardavan et~al.}(2007)\citenamefont{Ardavan, Ardavan,
  Singleton, Fasel, and Schmidt}}]{ArdavanH:Morph}
\bibinfo{author}{\bibfnamefont{H.}~\bibnamefont{Ardavan}},
  \bibinfo{author}{\bibfnamefont{A.}~\bibnamefont{Ardavan}},
  \bibinfo{author}{\bibfnamefont{J.}~\bibnamefont{Singleton}},
  \bibinfo{author}{\bibfnamefont{J.}~\bibnamefont{Fasel}}, \bibnamefont{and}
  \bibinfo{author}{\bibfnamefont{A.}~\bibnamefont{Schmidt}},
  \bibinfo{journal}{J. Opt. Soc. Am. A} \textbf{\bibinfo{volume}{24}},
  \bibinfo{pages}{2443} (\bibinfo{year}{2007}).

\bibitem[{\citenamefont{Borovikov}(1994)}]{BorvikovVA:UnStaPha}
\bibinfo{author}{\bibfnamefont{V.~A.} \bibnamefont{Borovikov}},
  \emph{\bibinfo{title}{Uniform Stationary Phase Method}}
  (\bibinfo{publisher}{Institution of Electrical Engineers},
  \bibinfo{address}{Stevenage, U.K}, \bibinfo{year}{1994}).

\bibitem[{\citenamefont{Jackson}(1999)}]{JacksonJD:Classical}
\bibinfo{author}{\bibfnamefont{J.~D.} \bibnamefont{Jackson}},
  \emph{\bibinfo{title}{Classical Electrodynamics}}
  (\bibinfo{publisher}{Wiley}, \bibinfo{address}{New York},
  \bibinfo{year}{1999}), \bibinfo{edition}{3rd} ed.

\bibitem[{\citenamefont{Popov et~al.}(2006)\citenamefont{Popov, Kuz'min,
  Ul'yanov, Deshpande, Ershov, Zakharenko, Kondrat'ev, Kostyuk, Losovskii, and
  Soglansnov}}]{PopovMV:Instant}
\bibinfo{author}{\bibfnamefont{A.~V.} \bibnamefont{Popov}},
  \bibinfo{author}{\bibfnamefont{A.~D.} \bibnamefont{Kuz'min}},
  \bibinfo{author}{\bibfnamefont{O.~M.} \bibnamefont{Ul'yanov}},
  \bibinfo{author}{\bibfnamefont{A.~A.} \bibnamefont{Deshpande}},
  \bibinfo{author}{\bibfnamefont{A.~A.} \bibnamefont{Ershov}},
  \bibinfo{author}{\bibfnamefont{V.~V.} \bibnamefont{Zakharenko}},
  \bibinfo{author}{\bibfnamefont{V.~I.} \bibnamefont{Kondrat'ev}},
  \bibinfo{author}{\bibfnamefont{S.~V.} \bibnamefont{Kostyuk}},
  \bibinfo{author}{\bibfnamefont{B.~Y.} \bibnamefont{Losovskii}},
  \bibnamefont{and} \bibinfo{author}{\bibfnamefont{V.~A.}
  \bibnamefont{Soglansnov}}, \bibinfo{journal}{Astron. Rep.}
  \textbf{\bibinfo{volume}{50}}, \bibinfo{pages}{562} (\bibinfo{year}{2006}),
  ISSN \bibinfo{issn}{1063-7729}.

\bibitem[{\citenamefont{Kinkhabwala and Thorsett}(2000)}]{KinkhabwalaA:Multi}
\bibinfo{author}{\bibfnamefont{A.}~\bibnamefont{Kinkhabwala}} \bibnamefont{and}
  \bibinfo{author}{\bibfnamefont{S.~E.} \bibnamefont{Thorsett}},
  \bibinfo{journal}{Astrophys. J.} \textbf{\bibinfo{volume}{535}},
  \bibinfo{pages}{365} (\bibinfo{year}{2000}).

\bibitem[{\citenamefont{Sallmen et~al.}(1999)\citenamefont{Sallmen, Backer,
  Hankins, Moffett, and Lundgren}}]{SallmenS:Simdog}
\bibinfo{author}{\bibfnamefont{S.}~\bibnamefont{Sallmen}},
  \bibinfo{author}{\bibfnamefont{D.~C.} \bibnamefont{Backer}},
  \bibinfo{author}{\bibfnamefont{T.~H.} \bibnamefont{Hankins}},
  \bibinfo{author}{\bibfnamefont{D.}~\bibnamefont{Moffett}}, \bibnamefont{and}
  \bibinfo{author}{\bibfnamefont{S.}~\bibnamefont{Lundgren}},
  \bibinfo{journal}{Astrophys. J.} \textbf{\bibinfo{volume}{517}},
  \bibinfo{pages}{460} (\bibinfo{year}{1999}), ISSN \bibinfo{issn}{0004-637X}.

\bibitem[{\citenamefont{Ardavan et~al.}(2003)\citenamefont{Ardavan, Ardavan,
  and Singleton}}]{ArdavanH:Fresfb}
\bibinfo{author}{\bibfnamefont{H.}~\bibnamefont{Ardavan}},
  \bibinfo{author}{\bibfnamefont{A.}~\bibnamefont{Ardavan}}, \bibnamefont{and}
  \bibinfo{author}{\bibfnamefont{J.}~\bibnamefont{Singleton}},
  \bibinfo{journal}{J. Opt. Soc. Am. A} \textbf{\bibinfo{volume}{20}},
  \bibinfo{pages}{2137} (\bibinfo{year}{2003}), ISSN \bibinfo{issn}{0740-3232}.

\bibitem[{\citenamefont{Lorimer and Kramer}(2005)}]{LorimerD:HbPA}
\bibinfo{author}{\bibfnamefont{D.}~\bibnamefont{Lorimer}} \bibnamefont{and}
  \bibinfo{author}{\bibfnamefont{M.}~\bibnamefont{Kramer}},
  \emph{\bibinfo{title}{Handbook of Pulsar Astronomy}}
  (\bibinfo{publisher}{Cambridge U. Press}, \bibinfo{address}{Cambridge, U.K},
  \bibinfo{year}{2005}).

\end{thebibliography}

\end{document}